\documentclass{iau}
\usepackage{graphicx}
\usepackage{natbib}
\usepackage{aas_macros}

\newcommand{\kms}{$\,$km$\,$s$^{-1}$}
\newcommand{\WHz}{$\,$W$\,$Hz$^{-1}$}
\newcommand{\ergs}{$\,$erg$\,$s$^{-1}$}

\newcommand{\msun}{{$M_\odot$}}
\newcommand{\msunyr}{{$M_\odot$ yr$^{-1}$}}

\def\HI{H{\,\small I}}

\def\emph#1{{\sl #1}}
\newcommand{\ltsima} {$\; \buildrel < \over \sim \;$}
\newcommand{\gtsima} {$\; \buildrel > \over \sim \;$}
\newcommand{\lta} {\lower.5ex\hbox{\ltsima}}
\newcommand{\gta} {\lower.5ex\hbox{\gtsima}}

\title[Radio jets: properties, life and impact] 
{Radio jets: properties, life and impact
}

\author[Raffaella Morganti]   
{Raffaella Morganti}

\affiliation{ASTRON, the Netherlands Institute for Radio Astronomy, Oude Hoogeveensedijk 4, 7991 PD Dwingeloo, The Netherlands\\
Kapteyn Astronomical Institute, University of Groningen, P.O. Box 800,
9700 AV Groningen, The Netherlands \\ email: {\tt morganti@astron.nl} }

\pubyear{2019}
\volume{356}  
\jname{Nuclear Activity in Galaxies Across Cosmic Time
Proceedings IAU Symposium No. 356, 2019}
\editors{M. Povi\'c et al.  eds.}
\begin{document}

\maketitle

\begin{abstract}
Our view of the properties of extragalactic radio jets and the impact they have on the host galaxy has expanded in the recent years. This has been possible thanks to the data from new or upgraded radio telescopes. 
This review briefly summarises the current status of the field and describes some of the exciting recent results and the surprises they have brought. In particular, the physical properties of radio jets as function of their radio power will be discussed together with the advance made in understanding the life-cycle of radio sources.  The evolutionary stage (e.g. newly born, dying, restarted) of the radio AGN can be derived from their morphology and properties of the radio spectra. The possibilities offered by the new generation of low-frequency radio telescopes make it possible to derive (at least to first order) the time-scale spent in each phase. The presence of a cycle of activity ensures a recurrent impact of the radio jets on their surrounding inter-stellar and inter-galactic medium and, therefore, their relevance for AGN feedback. 
The last part is dedicated to the recent results showing the effect of jets on the surrounding galactic medium. The predictions made by numerical simulations on the impact of a radio jet (and in particular a newly born jet) on a clumpy medium describe well what is seen by the observations.  The high resolution studies of jet-driven outflows of cold  gas  (\HI\ and molecular) has provided new important addition both in term of quantifying the impact of the outflows and their relevance for feedback as well as for providing an unexpected view of the physical conditions of the gas under these extreme conditions.  


\end{abstract}

\firstsection 
\section{Introduction}
\label{sec:intro}

\begin{figure}
\begin{center}
  \includegraphics[width=13cm]{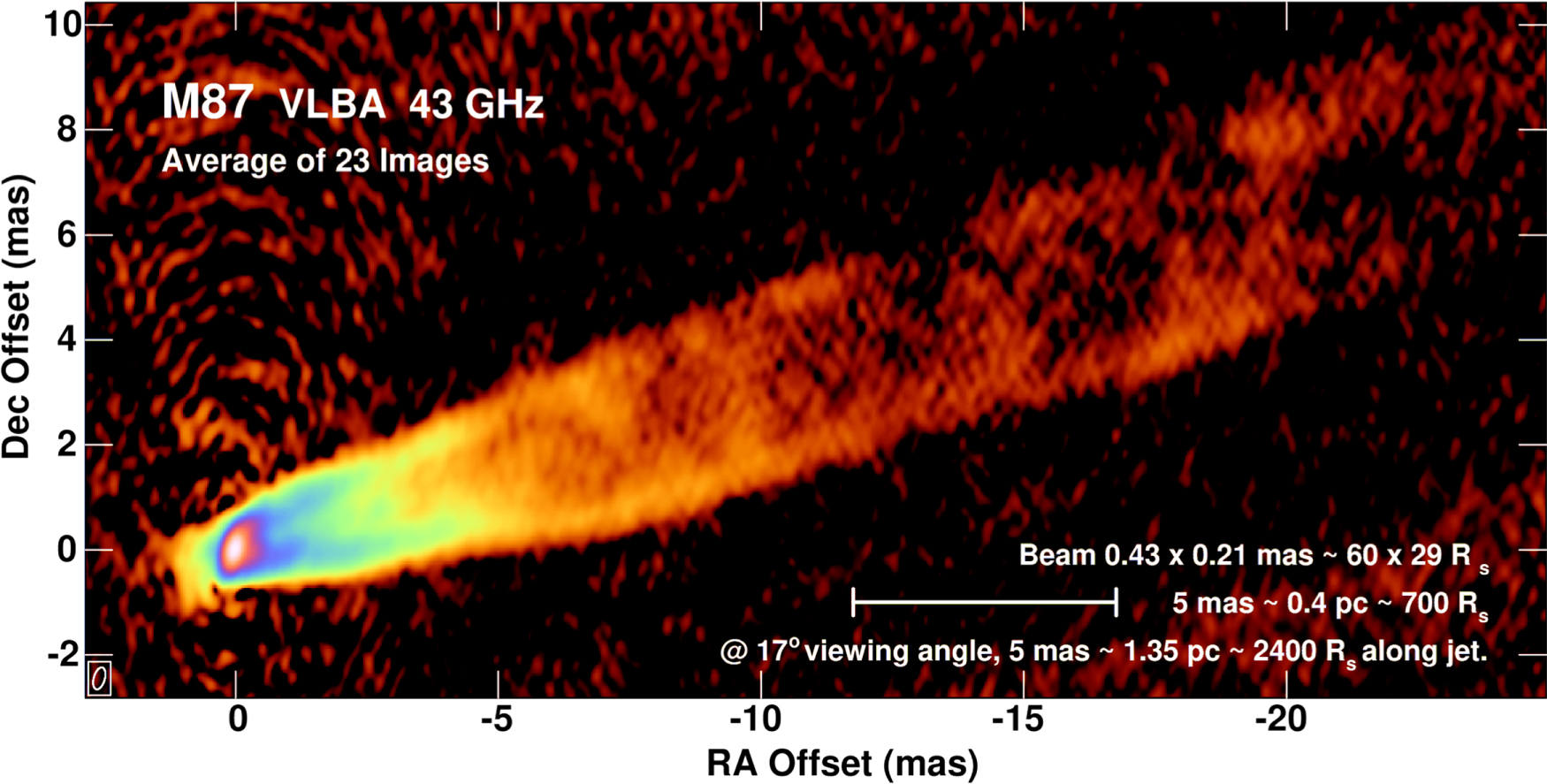}
 \caption{Very Long Baseline Array (VLBA) 43 GHz, 23-epoch average radio image of the jet and counterjet in M87 based on data from 2007 and 2008. Image taken from \cite{Walker18}; \copyright\  2018 The American Astronomical Society.}
\label{fig:M87}
\end{center}
\end{figure}

Radio jets are one spectacular manifestation of nuclear activity in galaxies. 
They are launched by a super massive black hole (SBMH) and collimated very close to it thanks to the combined effects of black hole spin and magnetic field \citep[see][for a review and Fig. \ref{fig:M87} for an example]{Blandford19}. 
This short review will discuss some of the latest results on their energetics, life-cycle and impact on the surrounding medium.  

Extragalactic radio jets span over a huge range of sizes (from pc to Mpc) and  show  a variety of structures and physical properties (see Sec. \ref{sec:powerjets}). They are also known to be recurrent during the life of a massive galaxy (see Sec. \ref{sec:LifeCycle}). All these properties have implications for the impact the jets have on the surrounding medium.  

Especially on cluster scale, they have been identified as responsible for preventing gas from cooling, therefore representing one of the clearest case of feedback in action driven by Active Galactic Nuclei (AGN)  \citep[see e.g. review by][]{McNamara12}.  However, their impact can also be visible on galaxy scales (see Sec. \ref{sec:Impact}) and can be considered complementary  to the effects of radiation and  AGN-driven winds.  

New results improving our understanding of radio jets and of their impact, have been made possible thanks to the upgrade of a number of  radio telescopes, like  JVLA, GMRT, WSRT-Apertif, eVLBI and  the coming in operation of new radio telescopes, like the Low Frequency Array \citep[LOFAR;][]{Haarlem13}, the Murchison Widefield Array \citep[MWA;][]{Tingay13} and ALMA. More radio telescopes are starting their operation, e.g. MeerKAT \citep{Jonas16}, and ASKAP \citep{Johnston07} or are planned for the coming years, i.e. SKA and SKA-VLBI \citep{Paragi15}, the latter particularly relevant for the connection to many countries in Africa. Thus, these are extremely interesting times for the study of radio jets.  Especially  relevant are the increased sensitivity and increased field-of-view that some of these telescopes have brought. Also important are the new possibilities offered by the low frequencies surveys, now reaching spatial resolutions comparable to what has been achieved so far only at GHz wavelengths. This is making possible the study of jets in this previously unexplored window. 
Particularly interesting are the results obtained by the LOFAR Two Meter Survey (LoTSS): an overview can be found in a special issue of A\&A (February 2019) and in  \cite{Shimwell19}. 

It is worth noting that the occurrence of radio emission and of radio jets is a strong function of the mass of the host galaxy and of the radio power of the source, see \cite{Sabater19} and refs therein.  Thus, in massive galaxies (M$_*  \gta 10^{11}$ \msun) the presence of a radio AGN with radio power $<10^{23}$ \WHz can reach above 80\%, while the fraction remains at most 30\% for radio sources above $10^{24}$ \WHz \citep{Sabater19}.

\begin{figure}
\begin{center}
  \includegraphics[width=13cm]{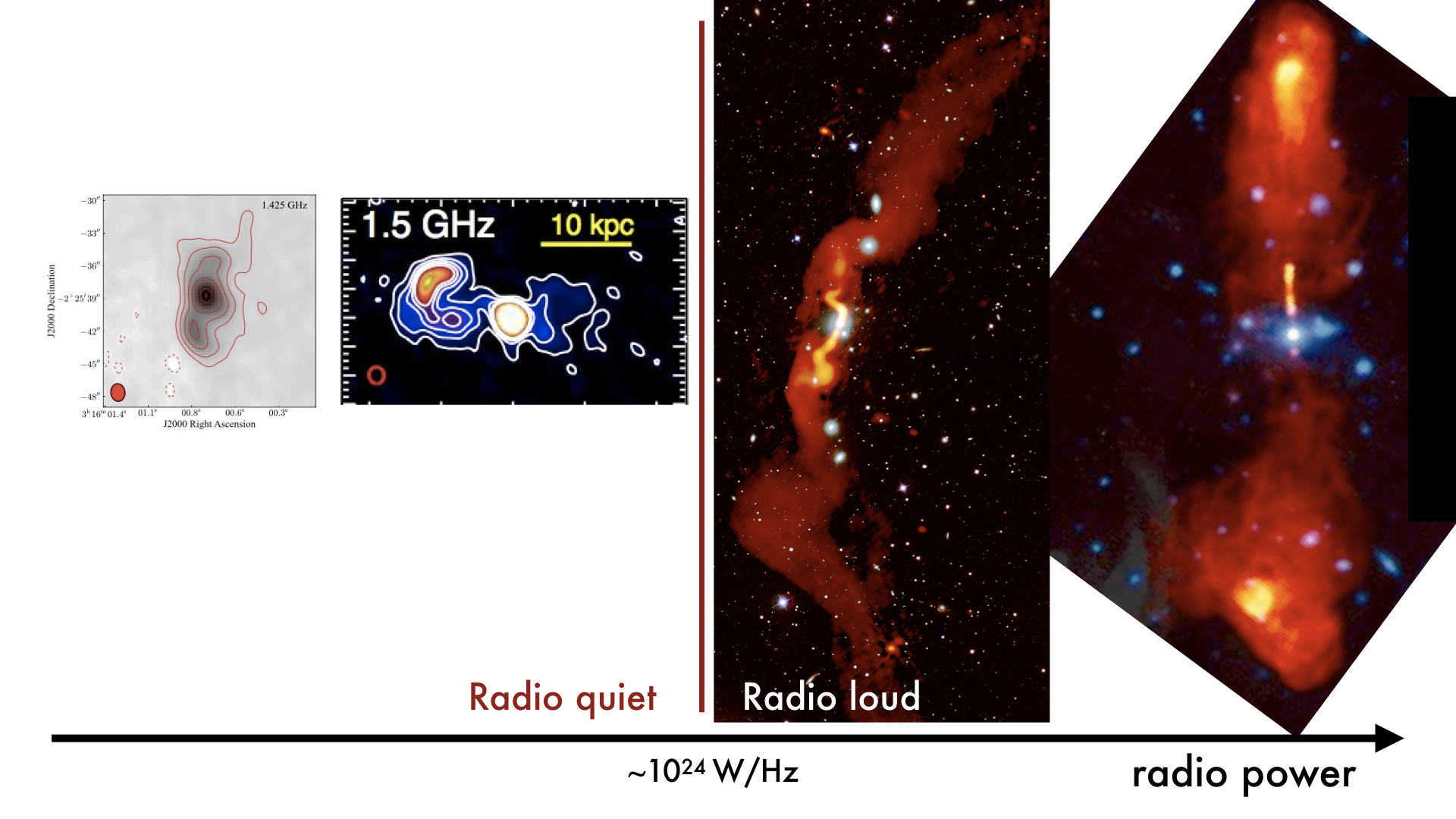}
 \caption{ Examples of morphology of radio jets observed in sources with different radio power. On the left two examples of radio-quiet sources.   NGC~1266 with radio power P$_{\rm 1.4~GHz} = 9.3 \times 10^{20}$ \WHz \citep{Alatalo11}, and the so-called "Tea cup" source with radio power P$_{\rm 1.4~GHz} = 5 \times 10^{23}$ \WHz  \citep{Harrison15} representing radio-quiet sources. On the right, two examples of radio loud sources. The giant FRI radio galaxy  3C~31 as observed by LOFAR at 150 MHz, from \cite{Heesen18} and  the FRII radio galaxy 3C~219 from \cite{Clarke92}  and Bridle Picture gallery; \copyright\  NRAO/AUI 1999.}
\label{fig:Morphology}
\end{center}
\end{figure}

\section{Jets as function of the radio power}
\label{sec:powerjets}

Figure \ref{fig:Morphology} shows examples of radio jets observed in  sources of different radio power. The figure illustrates the first-order dependence  of the morphology of the radio jet on this parameter. In the most powerful sources (P$_{\rm 1.4~GHz} \gta10^{24}$ \WHz), typically hosted by early-type galaxies, the jets can reach many hundred of kpc to above Mpc in size, something that does not happen among the low power radio sources.
The lower power sources (with power P$_{\rm 1.4~GHz} \lta 10^{24}$ \WHz), are often referred to as {\sl radio-quiet}  \citep[see][and discussion below]{Kellerman89}. 

Following the seminal work of \cite{Fanaroff74}, the radio-loud group of sources has been historically separated into two morphological classes, the  so-called Fanaroff-Riley I and II (FRI and FRII). The separation is considered to be a function of radio power, with FRII becoming prominent for power above P$_{\rm 1.4~GHz} \sim 10^{25.5}$ \WHz. However, a sharp separation in term of radio power has been recently questioned by \cite{Mingo19}, as result of the automatic classification performed on a large sample of radio galaxies produced by the LOFAR surveys. 
The Fanaroff-Riley division dependents on the properties of the jets and is, therefore, connected to the properties of the central engine (e.g. accretion rate, BH spin etc.) and to the environment into which the jet is expanding. Differences in the physical parameters of jets have been suggested and observed for FRI and FRII, e.g.  differences in the composition (i.e. fraction of thermal component), speed of the jet and spectral properties.
In particular, the spectral properties represent a powerful tool to trace the ageing of the relativistic electrons and test for its acceleration. 

For FRI, a detailed description of the observed properties can be found in \cite{Laing12,Laing14,Laing15} and refs therein. 
In general, the jets in these sources widen rapidly. They decelerate from relativistic to sub-relativistic speeds on scales of 1-10 kpc  \citep{Laing12}. The deceleration and the turbulence in the flow result in strong entrainment of external thermal medium. This helps balancing the pressure between the jet and the external medium, preventing the jet to be under-pressured, as otherwise suggested by X-ray observations \citep{Morganti87,Worrall00}. Thus,  jets in FRI sources can contain a large fraction of non-radiating, thermal medium.

The situation is different in FRII.  Hot spots are observed at the end of the jets indicating high Mach number jets and overpressure with respect to the medium. The spectral properties confirm this \citep{Harwood16}. In these jets, no strong entrainment is expected \citep{Croston18}.

Although according to the work of \cite{Sabater19} and refs therein, the stellar mass of the host galaxy (M$_*$) appears to have a stronger connection to the radio emission than the mass of the central BH (M$_{\rm BH}$),    
the link between the radio emission and the central engine is suggested by the properties of the optical emission lines of the host galaxy. In radio-loud galaxies the strength of these lines correlates, to first order, with the power of the radio source, i.e. the host of FRI radio galaxies show only weak emission lines while FRII tend to show strong emission lines  \citep[see][for a review on the optical properties of radio galaxies]{Tadhunter16}. The exception to this trend is represented by a group of FRII showing only weak emission lines, i.e. weak-line radio galaxies (WLRG). The nature of this group is still unclear \citep{Tadhunter16}. 

Despite the fact that most of the attention is given to extended sources, the great majority of radio sources are actually small ($< 20$ kpc) and appear unresolved in most of the surveys. Interestingly, in the past years more attention  has been given to these small sources. Indeed, even a new class in the Fanaroff-Riley classification has been introduced (not without controversies): the FR0 group \citep{Sadler14,Baldi15}. The nature of these radio sources and the role they play is still a matter of discussion \citep{Baldi19}. However, their relevance is that they release their energy on galaxy-scale. 

The so-called {\sl radio-quiet} sources (following Kellerman et al. 1989, these are sources with a relatively low radio-to-optical flux density ratios, $R<10$, and low radio powers, below $10^{24}$\WHz), represent a mix bag of objects, that includes Seyfert galaxies, quasars and low luminosity AGN (LLAGN). 
The definition of radio-quiet can be misleading and it should be kept in mind that these sources are not {\sl radio-silent} and in many cases they show radio jets \citep[see also discussion in][]{Padovani17}. 

Although as the radio power decreases it is more difficult to separate radio emission from AGN and starformation, a large number of radio-quiet sources have jets and should be treated as proper radio AGN \citep[see examples in samples studied by][]{Gallimore99,Morganti99,Jarvis19}. 
Despite the diversity observed in their host galaxy, 
all radio sources classified as radio-quiet have a number of properties in common. For example, they are typically of small sizes (up to at most a few tens of kpc), they have more complex morphologies and they are 
dominated by entrainment (due to the low velocity of their jets), therefore their jets can have a large fraction of thermal component.

Interestingly, the most recent work on a sample of radio-quiet, obscured quasars by \cite{Jarvis19} has shown that  the majority of the targets exhibit extended radio structures on 1 to 25 kpc scales. These radio features are associated with morphologically and kinematically distinct components in the ionised gas.
This is also confirmed by detailed studies of single, radio-quiet objects e.g. Seyfert galaxies like NGC~1068, IC~5063, Mrk 6 \citep[see][and refs therein]{Burillo14,Morganti15,Kharb14} and LLAGN \citep{Alatalo11,Combes13,Riffel14,Rodriguez17,Fabbiano18,Maksym19,Murthy19,Husemann19a}. 
Thus, the radio jets observed in these objects, despite their low power, can have an impact on the surrounding gas, disturb the kinematics and, in some cases, influence the physical properties (e.g. ionisation via shocks). This is further discussed in Sec \ref{sec:Impact}.

Regardless the morphology and classification of the radio jets, one of the parameters that is key for quantifying their impact on the surrounding interstellar medium (ISM) is the {\sl jet power}.  This gives the actual energetics available to the jet, combining radiative and non-radiative components, i.e. taking into account the thermal component.   Because of this, measuring the jet power is not trivial. 
Although other methods have also been proposed, see e.g. \cite{Willott99}, the studies of the X-ray cavities have provided a powerful way to estimate the jet power. By measuring the work needed to inflate the cavities, the power required from the radio plasma can be derived and then related to the radio luminosity \citep[see][for a review]{Cavagnolo10,McNamara12}.
The relation  derived in \cite{Cavagnolo10}, P$_{jet} \approx 5.8 \times 10^{43} (P_{radio}/10^{40})^{0.70}$ \ergs, is often used for this purpose. However, it is important to keep in mind the limitations of this approach, see discussion in \cite{Croston18,Shabala13}. 
In low power jets, the jet power is more uncertain because of the large fraction of thermal component coming from the entrainment. Because of this, the power of these jets can be larger than expected from their radio luminosity, see \cite{Bicknell98} and \cite{Mukherjee18a} for a recent example. Thus, given that these jets release their energy on galactic scales, they can play a relevant role in affecting the medium of their host galaxy and a better understanding of their jet power is important (see below).

\begin{figure}
\begin{center}
  \includegraphics[width=6cm]{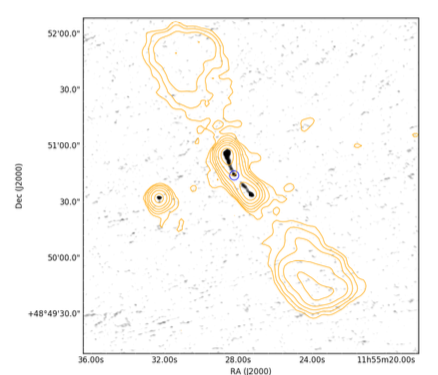} 
  \includegraphics[width=6cm]{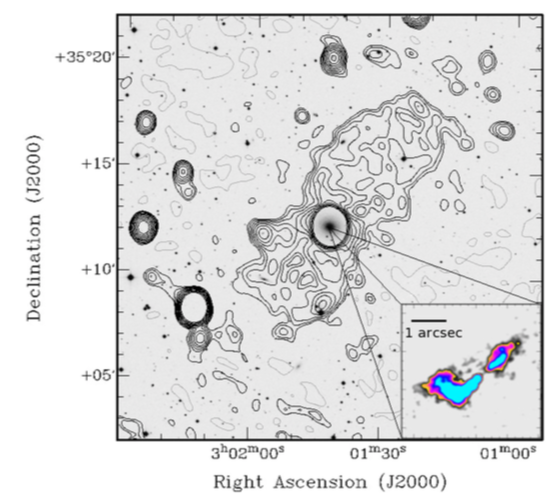} 
 \caption{Examples of restarted radio galaxies. {\sl Left:} double-double radio galaxy from the study of \cite{Mahatma19}. The image shows the emission at 1.4~GHz from the in greyscale, overlaid with the 144~MHz LOFAR contours from the LoTSS DR1 in orange. {\sl Right:} Contours of the diffuse radio emission around B2~0258+35 over-laid on a DSS2 image. Figure taken from \cite{Shulevski12}. 
 The inset at the bottom right shows the young, restarted source \citep[image taken from][]{Giroletti05}. A full discussion of the properties of this radiogalaxy can be found in  \cite{Brienza18}.}
   \label{fig:Restarted}
\end{center}
\end{figure}

\section{Life-cycle of radio jets}
\label{sec:LifeCycle}

The recurrence of the active phase of the super massive BH (SMBH) is a key ingredient in cosmological simulations of galaxy evolution, because it ensures that the energy released by the SMBH impacts the host galaxy multiple times as needed in the feedback cycle \citep[see e.g.][]{Ciotti10,Novak11,Gaspari17}. For radio AGN, we know that the period of activity is followed by a remnant phase when the nuclear activity switches off or drastically decreases. We also know that this cycle can repeat. The availability of deep (and high spatial resolution) surveys at low frequencies is a major steps forward for quantifying the time-scale of their cycle of activity \citep[see also][for an overview]{Morganti17a}.  This is because the emission at low frequencies remains, for the longest time, unaffected by energy losses, thereby acting as a "fossil record".

In radio AGN, we can identify newly started jets by the combination of small size and peaked radio spectra. Because of these properties, these young sources are usually called Compact Steep Spectrum (CSS) and GigaHertz Spectrum (GPS), see \cite{ODea98,Orienti16} for overviews of their properties. Their ages are typically $< 10^6$ yrs, and they have not emerged yet from the galactic medium (i.e. their size is $\lta 10$ kpc).
Most of these sources \citep[albeit not all, see ][for exceptions]{Kunert10} are believed to expand and break out from the galactic ISM, and grow to  evolved radiogalaxies.
Most of the life of a radio galaxy is spent in this phase (lasting roughly  few $\times 10^7$ up to  a few $\times 10^8$ yrs). Modelling of this active phase, in particular for FRII radio galaxies, have been presented in a number of studies, starting with the seminal work of \cite{Scheuer74}, to more recent studies, e.g. \cite{Kaiser07,Hardcastle18} and refs therein.  After this active phase, the jet can switch off.
In the dying phase, known as {\sl remnant phase}, the radio emission quickly fades away due to radiative and adiabatic losses, see e.g. \cite{Parma07,Brienza17} and refs therein.
Interestingly, radio AGN can also have a {\sl restart phase}. The best examples of this are the so-called "double-double" radio galaxies. An example can be seen in Fig. \ref{fig:Restarted} (left) and more details can be found in \cite{Schoenmakers00,Konar12,Mahatma19} and refs therein.  In these sources two sets of lobes are observed, resulting from two well separate phases of activity. The remnant phase in these objects has a duration comparable or shorter than their active phase, ranging from a few Myr to a few tens of Myr \citep[e.g.][]{Konar12}. 
Other type of restarted radio galaxies are known, for example those showing a bright inner, newly born radio source embedded in a low surface brightness emission, reminiscent of a remnant structure, see  Fig. \ref{fig:Restarted} (right) for an example. In these sources, the time scales of the active and remnant phases can only be derived from a detailed analysis of the radio spectrum \citep[see e.g.][]{Brienza18}.
 
Thanks to the low frequency radio data which are coming on-line (and in particular the data from the LOFAR surveys), we are getting a better view of the life cycle of radio galaxies.  \cite{Brienza17} and \cite{Hardcastle18} have  suggested that remnant sources fade rapidly, thus most of the observed remnant radio galaxies are relatively young, with total ages between $5 \times 10^7$ and $10^8$ yr. The majority of the remnant sources would be observed soon after the switch-off of the radio source and they are expected to evolve quickly due to dynamic expansion. The restarted phase can also follow  shortly after, in a similar way as found for the group of double-double radio galaxies (Jurlin et al. 2019). 
It is interesting to note that these findings are consistent with what derived from the study of X-ray cavities in radio galaxies in clusters and groups, see e.g. \cite{Randall11,Vantyghem14}. 
A full overview of the life cycle can only be obtained by including these findings in the context of modelling  the evolution with respect to the luminosity function of radio galaxies (e.g. Shabala et al. 2019).

\begin{figure}
\begin{center}
  \includegraphics[width=13cm]{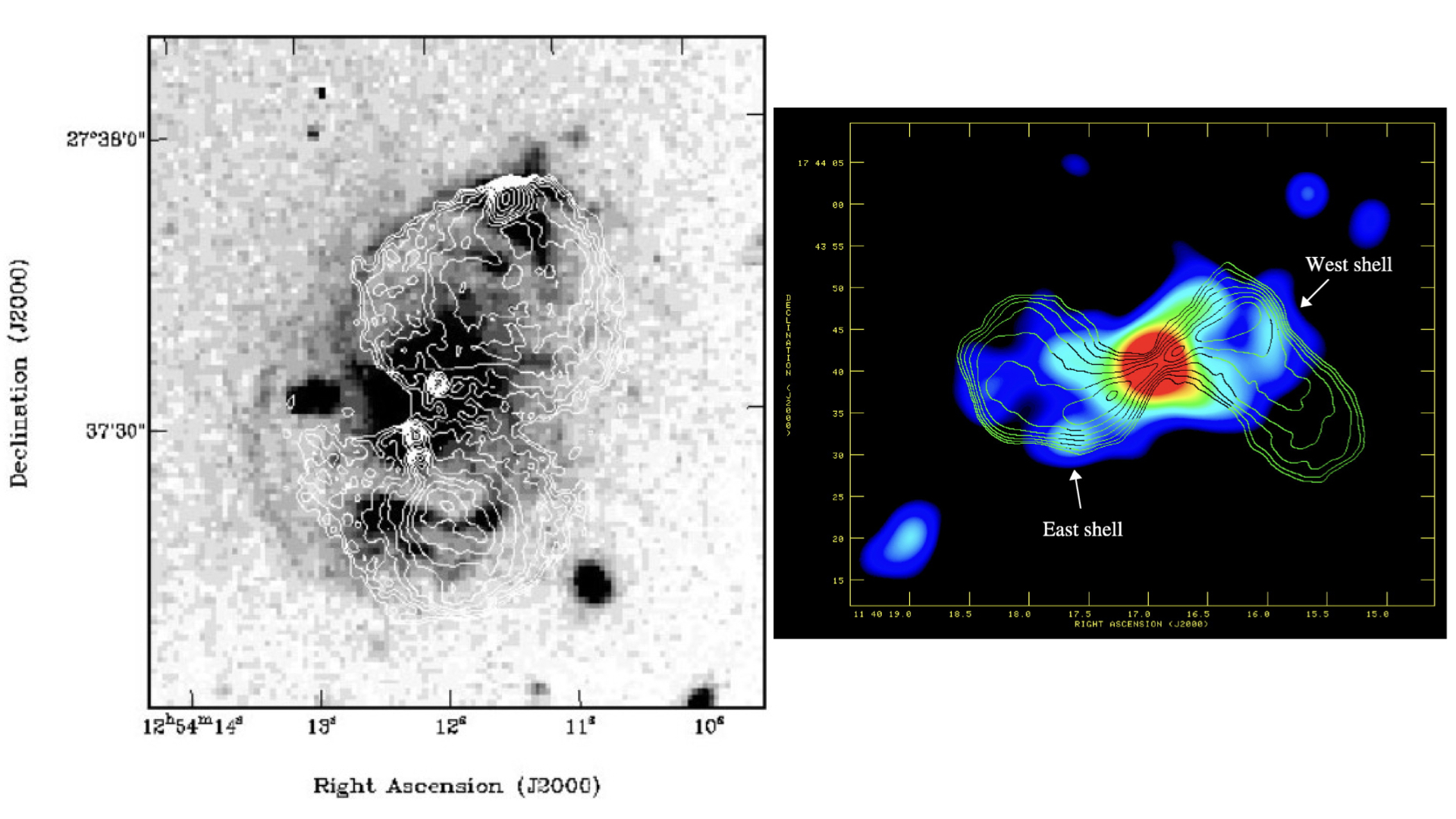} 
 \caption{ {\bf Left:} Overlay of the H$\alpha+$continuum image for Coma A (grey-scale) with the 6-cm radio map of \cite{Breugel85} (contours). Figure taken from \cite{Tadhunter00}. The  coincidence between the radio emission from the lobes and the ionised gas is clearly seen suggesting that the distribution of the gas is shaped by the interaction with the radio plasma. {\bf Right} Chandra image  with 1.4~GHz radio contours overlaid to illustrate the relationship between the X-ray shells and radio morphology in NGC~3801, figure taken from  \cite{Croston07}.}
   \label{fig:Impact}
\end{center}
\end{figure}

\section{Jets and their impact}
\label{sec:Impact}

Radio jets are recognised to play a role in feedback by preventing the cooling of hot (X-ray) gas surrounding central galaxies in clusters. Extensive work has been done on this, see e.g. \cite{McNamara12}. This role has been identified as {\sl jet-mode} feedback and associated to radio sources characterised by radiatively-inefficient accretion, i.e. FRI.  However, radio  jets can also drive massive gas outflows on galactic scales, an other signature of AGN feedback.  We will focus here on this role of the radio jets.  
An overview of  the relevance of outflows for feedback and galaxies evolution can also be found in a number of reviews, e.g. \cite{King15,Morganti17b,Harrison18} and refs therein. 

Radio jets have been known a long time  to be able to drive gas outflows \citep[see e.g.][]{Axon98,Capetti99}. Most of these studies focused on ionised gas (warm and hot). In the last years, however, the study of outflows has expanded also to the cold component of the gas: \HI\ and molecular \citep[see][as examples of earlier studies]{Morganti05a,Feruglio10}. 
The finding of cold gas associated with AGN-drive outflows has been quite surprising. The origin of this gas associated with fast outflows is still a matter of debate. The most likely explanation is that the component of cold gas is due to fast cooling after the gas is shocked by the interaction with the jet. 
Interestingly, in most cases studied so far, the cold phase of the gas appears to carry the larger mass of the outflow, i.e. higher compared to what associated to the warm ionised component of the outflow. 

In addition to this, the improvement of the numerical simulations describing the impact of jets on the surrounding medium has further helped in the intepretation of  the results from the observations. The new generation of numerical simulations assume more realistic conditions for the gas the jet expands into.  They are finding that radio jets couple strongly to the {\sl clumpy} ISM of the host galaxy, see \cite{Wagner12,Mukherjee16,Cielo18,Mukherjee18a,Mukherjee18b} for the details. According to these numerical simulations, a clumpy ISM, instead of a smooth one (as was assumed so  far in the simulations) is key in making the impact of the jet much larger than previously found: because of the clumpiness of the gaseous medium, the progress of the jet can be temporarily halted when hitting a dense gas cloud. Thus, the jet is meandering through the ISM to find the path of minimum resistance and, doing so, creating a cocoon of shocked gas driving an outflow in all directions \citep{Wagner12,Mukherjee16,Mukherjee18a}.
The jet power, the distribution of the surrounding medium and the orientation at which the jet enters the medium are all important parameters which determine the final impact of the jet-ISM interaction \citep{Mukherjee18a}.  

\subsection{Impact traced by the ionised gas}

As mentioned, the capability of radio jets to drive  outflows is known since long time thanks to the studies of the optical emission lines of the warm ionised gas and of X-ray emission tracing the presence of gas heated by shocks.
Evidence of jet-ISM interaction and jet-driven outflows have been found in a variety of objects. For example, Seyfert galaxies often show morphological association between the ionised gas and the radio emission. Furthermore, kinematical disturbance - traced by broad (often blueshifted) components of the emission lines - have been observed from  the gas co-spatial with the radio \citep[see e.g.][]{Capetti99,Axon98,Morganti07}. All these signatures  supports and highlight the role of the jets as responsible for the interaction.
Interestingly, also low-luminosity AGN show gas outflows attributed to radio jets as show in, e.g. \cite{Riffel14,Rodriguez17,May18} and refs therein.

In radio galaxies the presence of  kinematically disturbed gas at the location of the radio emission is a relatively common features. Figure \ref{fig:Impact} (left) shows the example of Coma~A and the spatial coincidence between the distribution of the ionised gas and the radio emission, strongly suggestive of a role of the radio plasma in shaping the distribution of the gas \citep{Tadhunter00}. The radio can also play a role in the ionisation of the gas but assessing this requires a detailed analysis of the line ratios and the comparison with shocks models. 
Particularly interesting is the result that young radio galaxies, i.e. sources hosting a newly formed jet, show these kinematically disturbance more often and of larger amplitude compared to large radio sources \citep{Holt08,Holt09}.  The effect of jets in high-z radio galaxies has been also studied in detail by e.g. \cite{Nesvadba08}. 

While outflows of ionised gas are very common in radio sources, the mass outflow rate associated with them is typically not high ($\lta 1$\msunyr), regardless whether they are driven by jets or by radiation/winds. Thus, the actual impact of these outflows for galaxy evolution is still an open question.  

Signatures of shocked-heated gas resulting from the interaction between the radio plasma and the ISM are also seen by X-ray observations. The case of NGC~3801 (Fig. \ref{fig:Impact}, right) nicely illustrates this process by showing shells of enhanced X-ray emission at the edge of  the radio lobes \citep{Croston07}.
It is interesting to note that this object is a low-power radio galaxy.  \cite{Croston07} estimate an expansion speed of the shells of  850 \kms, corresponding to a Mach number of 4 and this allows to measure directly the contribution of shock heating of the jet to the total energetic input to the ISM.  Similar conclusion about the relevance of shocks heating of the gas were derived for Centaurus~A \citep{Croston08}. In this object,  a shell of X-ray emitting gas is observed, tracing the effect of the southern radio lobe expanding with a velocity of $\sim 2600$\kms, roughly Mach 8 relative to the ambient medium. 

Signatures in the form of X-ray emitting gas heated by shocks from jet-ISM interaction have been reported also for the radio galaxy 3C~305, \citep{Hardcastle12}. 
The authors find that the X-ray emission is consistent with being shock-heated material and can be described by standard collisionally ionised models. Albeit with a number of assumptions, in this source the X-ray-emitting gas could dominate the other phases of the gas outflow \citep{Morganti05b} in terms of its energy content. This may be the case in more objects, but the lack of deep enough X-ray data prevent to draw strong conclusions.

\begin{figure}
\begin{center}
  \includegraphics[width=13cm]{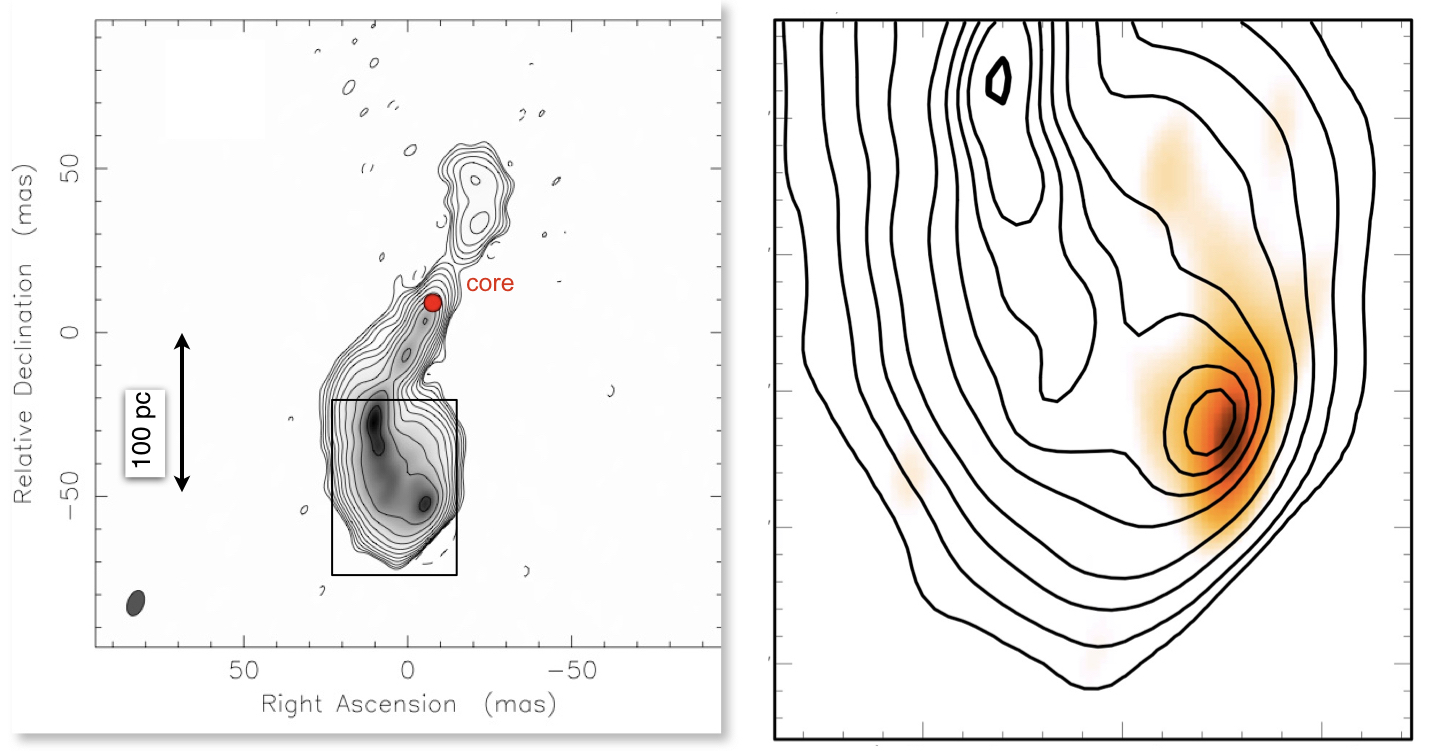} 
 \caption{{\bf Left:} radio continuum image of the young radio galaxy 4C~12.50. {\bf Right:} in orange is shown the distribution of the outflowing \HI\ (about 1000 \kms\ blueshifted compared to the systemic velocity). The spatial coincidence between the radio lobe and the gas, with a bright \HI\ cloud at the location of the hot spot and more diffuse gas wrapping around the lobe,  suggests that the outflow could be driven by the radio jet. 
Adapted from \cite{Morganti13}.}
   \label{fig:VLBI}
\end{center}
\end{figure}

\subsection{Jet-driven outflows of cold gas: \HI\ component}

A number of cases have been found where a component of \HI\ (identify by blueshifted  wings seen in \HI\ absorption) is associated with jet-driven outflows, see e.g. \cite{Morganti03,Morganti05a,Aditya18,Morganti18} and refs therein.  

In a shallow \HI\ survey of about 250 sources, at least 15\% of \HI\ detections show blueshifted components of \HI\ absorption \citep{Gereb15,Maccagni17}. Interestingly, the large majority of these outflows are observed in young (or restarted) radio galaxies \citep[see also][]{Aditya18}.  The \HI\ outflows can extend up to a few hundred pc to kpc in radius and
are characterised by velocities between a few hundred and $\sim 1300$ \kms, masses raging from a few $\times 10^6$ to $10^7$ \msun\ and mass outflow rates up to 20 - 50 \msunyr.
Their kinetic energies can be derived and compared to the Eddington (or bolometric) luminosities resulting in  $\dot{E}_{kin}/L_{edd} \sim 10^{-4}$ (few $\times  10^{-3}$ bolometric luminosity).

In a few cases, the \HI\ outflow has been located and spatially resolved using high resolution observations and VLBI. The best examples are 3C~293 \citep{Mahony16} and 4C~12.50 \citep{Morganti13}. Other examples can be found in \cite{Schulz18}. Figure \ref{fig:VLBI} shows the case of 4C~12.50, where the interaction jet-ISM is caught in the act. The outflowing \HI\ gas (shown in orange in Fig. \ref{fig:VLBI} right) is distributed around the head of the jet and on the side of the lobes, very suggestive of an on-going interaction.  From the VLBI observations it is also interesting to see that outflowing gas is  present very close to the AGN  \citep[$\sim 40$ pc,][]{Schulz18}. Furthermore, these studies also suggest that the structure of the outflow changes with time, with an increasing amount of diffuse gas - with respect to gas in clumps - as the radio source grows (Schulz et al. in prep). 

\subsection{Jet-driven outflows of cold gas: molecular  component}

The study of molecular gas in radio galaxies in a variety of environments is rapidly expanding, in particular thanks to the capabilities of ALMA \cite[see e.g.][]{Ruffa19,Russell19}. 

The sensitivity of ALMA has also made possible to derive  the best evidence of fast and massive outflows driven by radio jets. One of the objects where the effects of the radio jet on the molecular gas has been studied in details is the Seyfert 2 galaxy IC~5063. This object is also one of the clearest examples of low-power jet (this source would be classified as radio-quiet) disturbing the kinematics of {\sl all the phases of the gas} \citep[see][for an overview]{Tadhunter14,Morganti15}.

The ALMA observations have confirmed the presence of molecular gas with disturbed kinematics across the entire region co-spatial with the radio emission  \citep{Morganti15}. The observations of a number of CO transitions have shown that in the immediate vicinity of the radio jet, a fast outflow, with velocities up to 800 \kms, is occurring. In addition to this, the interaction is also affecting the  physical conditions of the gas  \citep{Dasyra16,Oosterloo17}. 
The gas involved in the outflow has high excitation  temperatures (in the range 30 - 55 K)  and, based on the relative brightness of the $^{12}$CO lines and of $^{13}$CO(2-1) vs $^{12}$CO(2-1), the outflow must be optically thin \citep[see][for details]{Oosterloo17}. The mass of the molecular outflow is estimated to be at least $1.2 \times 10^6$ \msun\ and the mass outflow rate is $\sim 12$\msunyr. Although not extremely high, it is much higher than the one derived for the warm ionised gas.  
Interestingly, the kinematics of the gas can be well reproduced by the  hydrodynamic simulations described above, which model the effect of the radio jets on the multiphase, clumpy interstellar medium. The detail of the simulation and the results of the comparison are described in \cite{Mukherjee18a}.  

\begin{figure}
\begin{center}
  \includegraphics[angle=0,width=13cm]{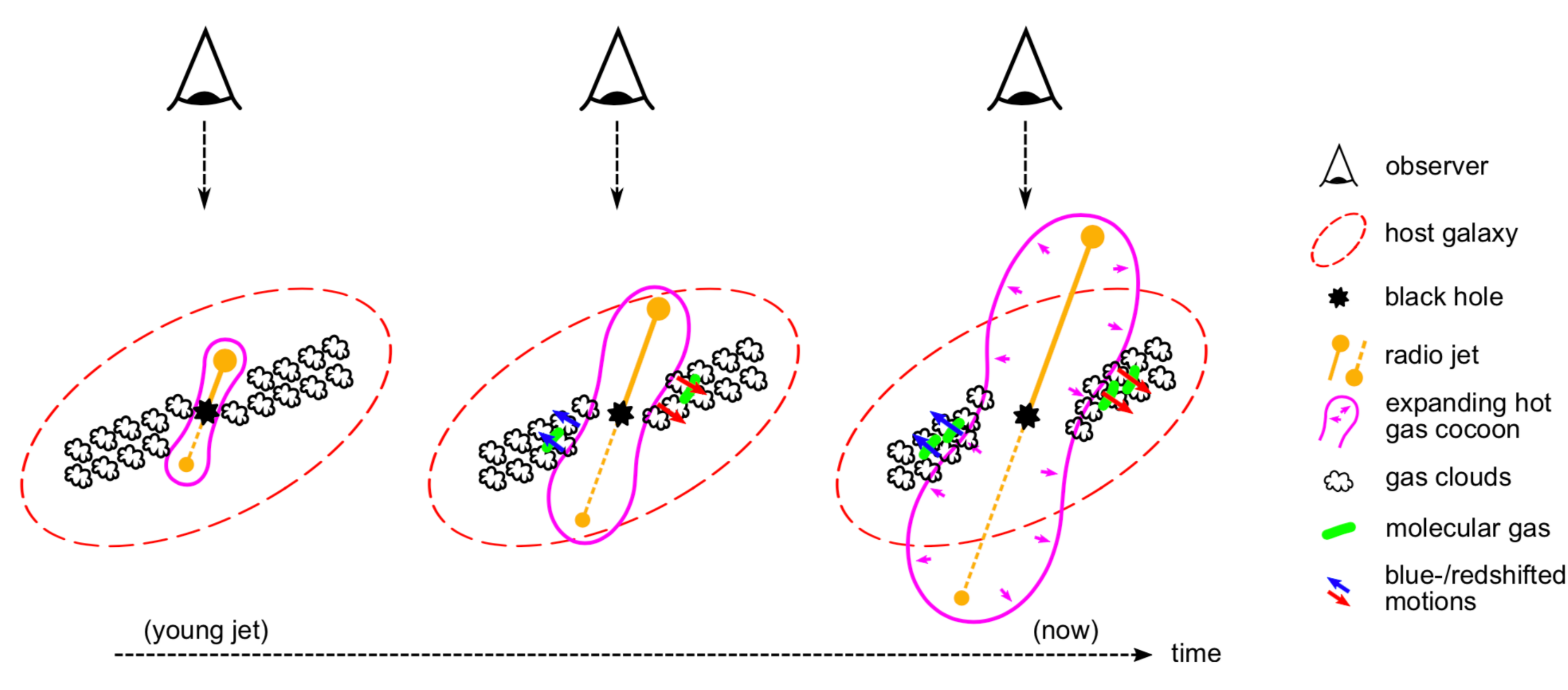} 
 \caption{Cartoon of the jet-ISM interaction in the case of 3C~273 as proposed by \cite{Husemann19b}. The sketch illustrates the dependence of the impact from the geometry of the jet-disk system.This has been suggested to occur also in other objects (see text) and is predicted by the numerical simulations of \cite{Mukherjee18b}. }
   \label{fig:3C273}
\end{center}
\end{figure}

In addition to IC~5063, in a steadily growing number of radio galaxies (both radio-quiet and radio-loud) the interaction of the jets with dense clumps producing molecular outflows has been observed. This is thanks not only to the depth but also to the high spatial resolution provided by ALMA. Some examples can be found in e.g.
NGC~1068 \citep{Burillo14}, HE~1353-1917 \citep{Husemann19a}, NGC~613 \citep{Audibert19}, 4C12.50 \citep{Fotopoulou19}, PKS1549-69 \citep{Oosterloo19}, 3C~273 \citep{Husemann19b}.
The actual impact of the jet-ISM interaction will depend on the distribution of the gas and the orientation of the jet compared to it. Interestingly, this dependence is seen both in low and high radio power jets.  This is seen very clearly in IC~5063 but it is supported by other cases as extensively discussed in \citep{Mukherjee18b}. Other examples are  e.g. the low-power AGN HE~1353-1917 \citep{Husemann19a} and the high-power jet in 3C~273 \citep{Husemann19b}. The cartoon shown in Fig. \ref{fig:3C273} illustrates the scenario proposed by the authors for this object but also representing what predicted by the numerical simulations: the radio jets is creating a  pressurised expanding hot gas cocoon which is impacting on the inclined gas disk. 

\begin{figure}
\begin{center}
  \includegraphics[width=13cm]{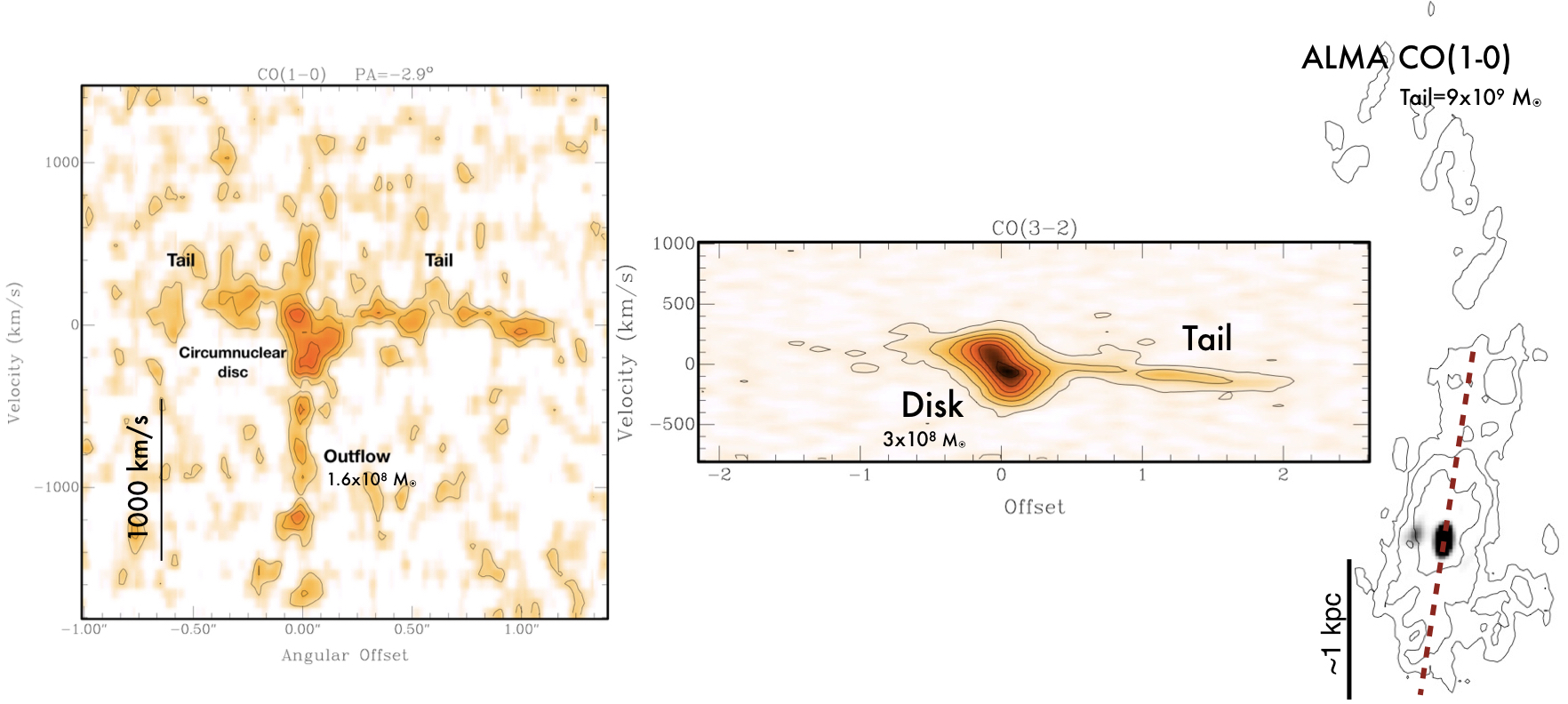} 
 \caption{The complex view of the central few hundred pc of PKS~1549-69 ($z=0.150$), see \cite{Oosterloo19} for details. ALMA CO(1-0) and CO(3-2) detected in emission with spatial resolution ranging from 0.05 arcsec ($\sim 100$ pc) to 0.2 arcsec.}
   \label{fig:pks1549}
\end{center}
\end{figure}

The presence of radio jets  does not preclude other forms of AGN activity (like radiation and winds) impacting the ISM. In some cases,  it may even be difficult to disentangle the effect of these different phenomena. 
The case of PKS~1549-69, an object hosting an obscured quasars and a young radio source,  is particularly interesting. 
The ALMA high resolution observations show the presence of  three gas structures, which can be seen in Fig. \ref{fig:pks1549}, tracing the accretion and the outflowing of molecular gas. Kiloparsec-scale tails are observed, resulting from an on-going merger and  providing gas which accretes  onto the centre of PKS~1549-79. At the same time, a circum-nuclear disc has formed in the inner few hundred parsec, and a very broad ($> 2300$ \kms) component associated with fast outflowing gas is detected in CO(1-0) at the position of the AGN. As expected, the outflow is massive (up to 600 \msunyr) but, despite the fact that PKS~1549-79 should represent an ideal case of feedback in action, it is limited to the inner 200 pc. 
Both the jet and the wind/radiation can drive the outflow (and perhaps they both do!). Only circumstantial evidence suggest the jet may play a prominent role, as the jet appears to be affect by strong interaction with the ISM, possibly providing the driving mechanism for the massive outflow.
These results illustrates that the impact on the surrounding medium of the energy released by the AGN is not always as expected from the feedback scenario.

\section{Conclusions}

In summary, radio jets show a variety of  structures, physical conditions (velocity, composition etc.) and energetics.
The importance of  radio jets goes beyond the radio-loud objects and the (much more common) radio-quiet sources should also be considered.  Their radio emission shows often jet-like structures and the power of these jets can have a relevant impact on the host galaxy. They can interact with the ISM of the host galaxy for a longer period and, integrated over time, they can dump large amounts of energy in the ISM.  

Radio jets are also a recurrent phenomenon.  From their properties (e.g. morphology and radio spectra) we can identify young, dying and restarted radio sources. The recurrence of the radio AGN ensures that they  
impact the host galaxy multiple times as needed in the feedback cycle. Low frequency radio surveys are now helping to derive more reliable time-scales of their life-cycle. 

Jets can drive (massive) multi-phase outflows. Based on new numerical simulations, the predictions are that jets couple well with clumpy ISM and that they can produce cocoons of shocked gas expanding across the surrounding gas. These effects are already observed in a growing number of objects. 
Young (or restarted) jets are those showing the most clear signs of affecting the surrounding medium. This is also consistent with the predictions from the simulations.

Thanks to the increasing sensitivity and the capabilities offered by the new radio telescopes, our understanding of both the physical properties as well as the impact of radio jets, is continuously expanding. These progresses, combined with the study of AGN at other wavelengths, will hopefully help building a more complete picture of the interplay between AGN and host galaxy, a problem that still has many open questions.

\acknowledgements{I would like to thanks the organisers and, in particular, Mirjana  Povi\'c for putting together such an interesting (not only scientifically!) conference. Some of the  results presented here would not have been obtained without the help of some key  collaborators. In particular I would like to thank Tom Oosterloo, Robert Schulz and Clive Tadhunter}.

\end{document}